# Speckle reduction in ocular wave-front sensing

Vassilios Albanis[1], Erez N. Ribak[1] and Yuval Carmon[1,2]

[1]*Department of Physics, Technion-Israel Institute of Technology, Technion City, Haifa 32000 Israel*
*valbanis@physics.technion.ac.il, eribak@physics.technion.ac.il,*

[2]*Shamir Optical Industry, Kibbutz Shamir, Upper Galilee 12135, Israel*
*yuval@shamir.co.il*

**Abstract:** An acousto-optic cell was used to reduce the speckle noise that reduces the quality of Hartmann-Shack and other wave-front sensors measuring ocular aberrations. In the method presented here, a laser beam traverses an acousto-optic cell, interacting with standing acoustic waves. Speckle reduction takes place as the incoming beam is diffractively spread across the cornea. The increased size and the wider angular spread of the incoming beam average out the speckles, producing a more uniform response of the wave-front sensor.

**References and links**

1. A. V. Larichev, P. V. Ivanov, I. G. Iroshnikov, V. I. Shmal'gauzen, "Measurement of eye aberrations in a speckle field", Quantum Electronics (Kvantovaya Elektronika) **31,** 1108-1112 (2001).
2. H. Hofer, P. Artal, B. Singer, J. L. Aragon and D. R. Williams, "Dynamics of the eye's wave aberration", J. Opt. Soc. Am. A **18**,597–606 (2001).
3. J. Rha, R. S. Jonnal, K. E. Thorn, J. Qu, Y. Zhang and D. T. Miller, "Adaptive optics flood-illumination camera for high speed retinal imaging", Opt. Express **14**, 4552-4569 (2006).
4. J. Selb, S. Leveque-Fort, L. Pottier, A. C. Boccara, "3D acousto-optic modulated-speckle imaging in biological tissues", C. R. Acad. Sci. Paris, t. **2**, Série IV, p. 1213–1225, (2001)
5. D. P. Popescu, M. D. Hewko, M.G. Sowa, "Speckle noise attenuation in optical coherence tomography by compounding images acquired at different positions of the sample', Optics Communications **269**, 247-251 (2001).
6. E. N. Ribak, "Harnessing caustics for wave front sensing", Optics Letters **26**, 1834-1836 (2001).
7. V. Albanis, Y. Benny, Y. Carmon and E. N. Ribak, "Ocular wavefront sensing using an acousto-optic cell as an adjustable Hartmann-Shack sensor", 5th Aegean Summer School in Visual Optics, Rethymnon Crete, Ed. Harilaos Ginis, Institute of Vision and Optics University of Crete, Heraklion, Greece, July 1-6 (2006).
8. Y. Carmon and E. N. Ribak, "Phase retrieval by demodulation of a Hartmann-Shack sensor", Optics Communications **215,** 285 - 288 (2003).
9. Y. Carmon and E. N. Ribak, "Fast Fourier demodulation", Applied Physics Letters **84,** 4656-7 (2004).
10. A Talmi and E N Ribak, "Direct demodulation of Hartmann-Shack patterns", J. Opt. Soc. Am. A **21,** 632-9 (2004).
11. C. Canovas and E. N. Ribak, "Comparison of Hartmann analysis methods", Applied Optics **46,** 1830-5 (2007).
12. I. Iglesias, R. Ragazzoni,Y. Julien and P. Artal, "Extended source pyramid wave-front sensor for the human eye", Optics Express **10,** 419-428 (2002).
13. S. R. Chamot, C. Dainty and S. Esposito, "Adaptive optics for ophthalmic applications using a pyramid wavefront sensor", Optics Express **14,** 518-526 (2006).
14. C. Torti, L. Diaz-Santana, D. Cuevaz and P. Harrison, "Wavefront Sensing of Human eyes using Curvature-based Sensors" in Imaging in the eye III: Technologies and clinical applications*,* 22 September, London, UK (2006)

**1. Introduction**

Wave-front sensing by means of a Hartmann-Shack (HS) sensor is a well established technique and a proven method, readily used to examine the aberrations of optical systems. On the practical level, however, one of the most commonly encountered limitations is the presence of speckle noise that typically arises from the coherent nature of the light sources used, and their interaction with the biological tissue examined. In the case of ocular wave-front measurements, the retina creates most of the unwanted noise in the form of a strong speckle field in the reflected laser light, due to the random interference of the coherent light in the highly anisotropic tissues comprising the retina. The resulting nonlinear response and saturation in the sensor, and therefore the accuracy of the reconstructed wave-front, are strongly affected [1]. A number of different methods have been employed in an effort to reduce the speckle effect, such as scanning and descanning mirrors [1, 2], diffusers such as rotating scatter-plates, multi-mode or dispersive fibres [3]. Decreasing the coherence of the light also achieves this, such as by the use super-luminescent diodes (SLDs), which have a coherence length of 20 to 30 μm [3] or even less coherent femtosecond lasers as used in optical coherence tomography, and even white light. However, the collection efficiency of SLDs and fibred beams is low and requires more lenses and pinholes. To balance that, higher power SLDs can be used, but again they have narrower spectral width and thus longer coherence length. In other cases, the interaction of light and sound has been used to reduce the speckle noise employing photo-acoustic and acousto-optic methods [4], as well as compound imaging applied *in vitro* for biological tissue imaging as [5], which again are not practically suited for imaging of the human eye.

In this work we present a simple and complimentary method to remove the speckle noise during HS measurements of ocular aberrations, and increase the accuracy of the reconstructed wave-front. We average the speckle field by widening the beam slightly on the cornea, both in real space and in *k* space. This in turn scatters off a larger volume in the retina. Thus the random speckle effect is averaged out to a significant degree. The wider beam is obtained through interaction between a laser beam and standing acoustic waves. The device has been previously used to produce a HS sensor with a variable pitch, where the lenslet array is replaced by caustics in the interaction region between two standing acoustic waves perpendicularly to the incoming laser beam, as viewed in the near field [6,7]. In the method presented herein, we use an acoustic cell in a similar manner, with the distinction that it is operated outside the near field, as we detail below.

**2. Hartmann-Shack system**

We implemented our acousto-optic method with a minimal modification of the HS wave-front sensing arrangement, Fig. 1. A polarized laser diode, operating at 780 nm, was used to illuminate the subject eye, with a nominal beam diameter on the cornea of 1 mm or slightly less. The acoustic cell was placed right after the laser. Eight percent of the light was reflected, using a beam splitter, towards the eye, some 40 cm down the beam from the cell. The laser power never exceeded 60 μW, and we employed an analyzer to tune it down to 20-30 μW during measurements (this is about a tenth of the permitted exposure). The incident light met the eye about 1-3 mm off the optical axis and parallel to it, so as to avoid unnecessary reflections from the cornea. After hitting the retina the light was scattered out through the pupil, then used to determine the aberrations of the subject eye. For dilated pupils the unwanted off-axis corneal reflection was removed by an aperture placed conjugate to the retina. With narrower pupils, the corneal reflection was closer to the axis and had to be cut down by an analyzer. No effort was made to reduce the polarisation of the laser, in order to keep the system simple, even if the speckle was slightly worse. The illuminated pupil was imaged onto a glass microlens array, consisting of 1.9 mm focal length, 190 μm diameter lenslets. The multiple foci were relayed again to be imaged by a standard uncooled CCD camera. For denser sampling of the wave front we used approximately half a dozen lenslets per mm on the cornea.

Compared to the usual ocular wave front sensor, the Hartmann spots are not well separated (Section 4), but the analysis method that we use is not sensitive to this contrast (Section 5). As a result, the retinal spot was not resolved by these lenslets, even if extended to ~80 µm. Thus, because of the large number of lenslets used to sample the pupil, the shape and size of the retinal spot is has minimal consequence on our analysis method. Centroiding, on less dense arrays, will also be minimally affected by changes to the spot shape. An optometric corrector, incorporating a fixation target, was adjusted subjectively to compensate for the defocus of the subjects' eyes. In an additional optical arm of the setup, a second CCD camera was trained on the subject's retina in order to acquire images of the point spread function of the light beam.

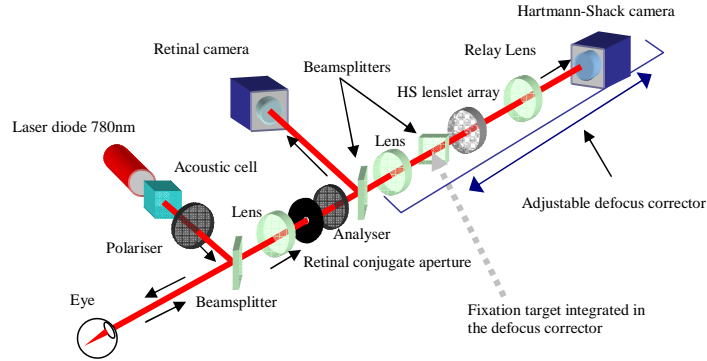

Fig. 1. The ocular wave front sensor setup used in the speckle reduction method. The acoustic cell is positioned immediately after the laser diode. The power of the incident beam is controlled by the polarizer. On reflection the retinal beam travels through the optical path, to the HS microlens array and it is then imaged on the HS camera. The PSF is also recorded at the retinal camera arm. An aperture and an analyzer can be optionally used to block any unwanted reflections from the subject's cornea.

## 3. The acoustic cell

The acoustic cell was positioned immediately after the laser diode with its $x$ and $y$ axes matching the horizontal and vertical planes of the optical setup, and the $z$ axis along the optical axis of the setup. It is of rectangular construction, $46 \times 53 \times 50$ mm in size, water filled, with two windows across the $z$ direction. The cell is extremely easy to construct and use: four flat piezoelectric drivers are glued to a thin-wall rectangular duct, and two windows are glued to seal the open sides [6]. A sine wave generator drives the two pairs of piezoelectric transducers, positioned at the four opposite sides of the cell. When operated, the transducer pairs create two sets of standing acoustic waves, at right angles to each other (Fig. 2), over a broad range of frequencies, with prominent resonances appearing at 1.2 MHZ, 3.5 MHz and 6 MHz, and a number of lesser ones in between. The laser light traverses the cell interacting with the two acoustic columns, formed by the standing waves, in the Raman-Nath regime. As a result, a number of positive and negative orders of diffraction are produced in the $x$ and $y$ directions. The number of orders, and hence the apparent size of the spot, are dependent upon and can be controlled by the proximity to the main resonant frequencies and the power supplied to the cell. The effect might be seen between resonances, but the power required would be higher.

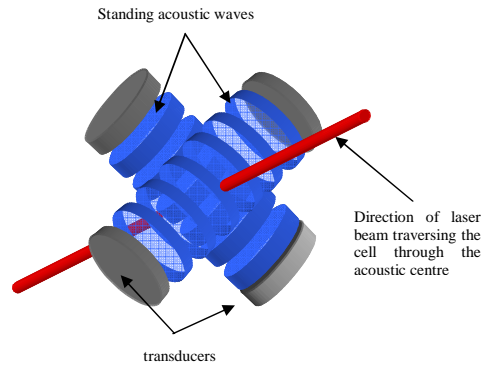

Fig. 2. The principle behind the operation of the acousto-optic method. Two sets of transducers operated at the same frequency will create two acoustic standing wave columns at right angles to each other. Light traversing perpendicularly through the acoustic centre interacts diffractively with the pressure waves, to create ±1, ±2,,, orders across each standing wave. The node spacing of the standing waves in water at few MHz is on the scale of a millimeter, comparable to the diameter of the laser beam.

At the lower range of resonant frequencies, below 1.5MHz, the laser spot appears diffractively elongated in a continuous manner, but at the higher frequencies the orders are distinct and clearly separated. By tuning the frequency, standing waves are formed in *x* and/or in *y*, as the two cavity dimensions (*x* and *y*) are not equal, thus creating high order beams in either or both directions. The apparent size of the beam on the cornea, when the acoustic cell is operated at different acoustic powers is of the order of 1-2 mm. If the eye is focused at infinity and the cell is ~40 cm away, the retina also observes and resolves the first and higher orders.

**4. Procedure**

To test the reduction of speckle noise using our acousto-optic method, we examined the aberrations of the left eyes of three experienced subjects with varying degrees of defocus and astigmatism. The age range was between 32 and 56 years old, with defocus in the range of +1.5D to -4.5D, and cylinder from -0.25D to -1.0D. The subjects were asked to fix their gaze on a target, or fixate with their other eye at a monitor viewing their own HS pattern, placed at infinity, on or within 2° of the optical axis. The pupils were naturally dilated and the images were acquired under darkness conditions, allowing for the subjects to become accustomed to the low light for few minutes before measurement. The pupil size was measured to be between 4 and 7 mm when the wave front sensing was taking place. The images of the HS patterns with the cell off and on were acquired consecutively while the subject was kept at the same position on a head-rest and gazed at the same point. Each image acquisition lasted approximately four seconds, during which twelve HS images were captured, each capture lasting 40 ms. Each image was used to reconstruct the wave-front separately, and the average image and series variance were calculated. The retinal camera was used to record the double-pass image of the diffracted spot on the subject's retina with the acoustic cell operating at different frequencies (Fig. 3). This provided us with an interesting situation in which an observer (or the subject himself) sees, at the monitor of the retinal camera, an image of the spot that closely resembles what the subject sees. We found out that the subjective and objective impressions are extremely similar.

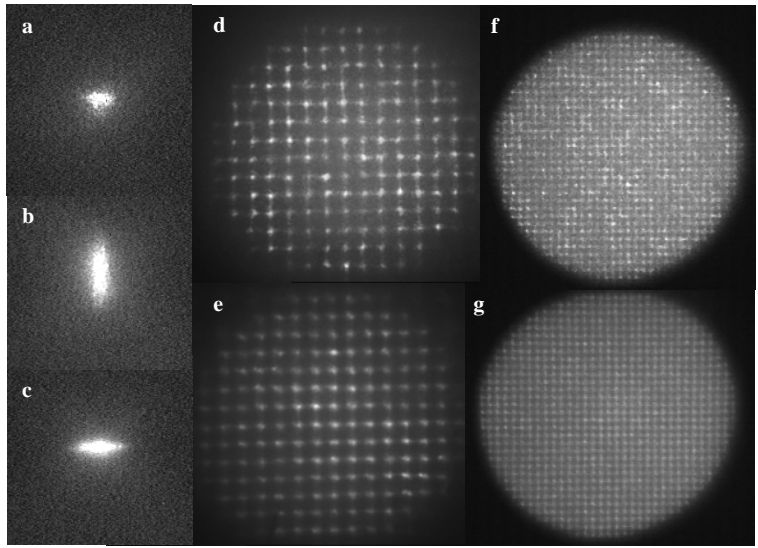

Fig. 3. Retinal and HS images as acquired with the acoustic cell on and off. (a) retinal spot with the cell off, (b) retinal image with the cell on and diffraction across the *y* direction, (c) retinal image with diffraction along the *x* direction. (d) pupil image acquired with the cell off showing the speckled and saturated HS pattern and (e) the same HS pattern with the cell on, illustrating the reduction in speckle; (f) pupil image sampled at much higher lenslet frequency with the cell off and (g) with the cell on.

## 5. Analysis

Using the HS images acquired with the acoustic cell in operation, we proceeded to reconstruct the associated wave-fronts. Usually one employs centroiding of the lenslet foci spots to find their shifts, corresponding to the wave front slopes. From these slopes the different Zernike coefficients are extracted. Instead we used the super-heterodyne method, where we demodulate the main lenslet frequency to yield the slopes [8] or directly the wave front [9], either in the Fourier domain or directly on the image [10]. In the Fourier analysis method, it is beneficial to have many HS spots so that the resolution and accuracy of the results is enhanced [8-10]. The high density of spots might lead to a pattern with reduced contrast, but with better resolution and somewhat better accuracy [11]. Zernike coefficients are derived from the final wave fronts. To validate the analysis, we used an artificial eye, where the retina, at the focus of a misaligned lens, was simulated by a moving white paper. The speckle properties were different from a live eye [1], but strong enough to saturate the detector, and thus to distort the Hartmann pattern. When we switched the cell on, saturation disappeared and we could inject more light into the artificial eye. To test the accuracy of the method we averaged 100 images taken with the laser, each with its speckle pattern, at very low laser light level so as not to saturate the detector. The result, with smooth but distinct Hartmann spots, was compared to a reference pattern taken with a white light source behind a pinhole at the lens focus (that is, a single-pass reference). The procedure in Section 4 was now performed with the cell turned off, and then resonating at two different frequencies. Here we also used the white light pattern as a reference, and calculated a sequence of wave fronts. Fig. 4 compares the Zernike coefficients until they become negligible, for the case of a single smooth pattern, and for the three cases of series of speckled images, with the cell off, at 1.25 MHz and at 3.30 Mhz. It is clear that with sufficient averaging, the results are quite similar. However, the advantage of the cell does not become evident until the comparison is performed under ocular scattering conditions [1], as is shown next.

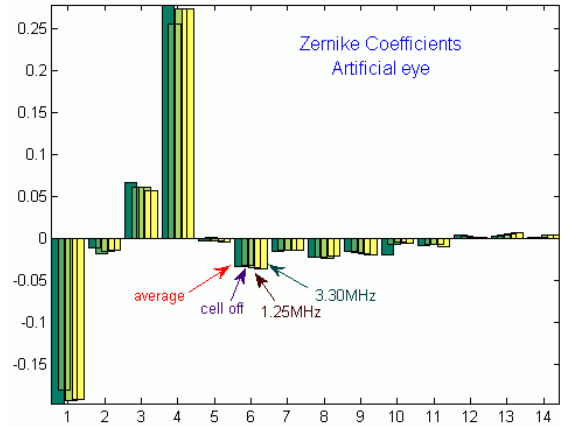

Fig. 4. Comparison of the first Zernike terms for an artificial eye. In each coefficient, bars (from the left) represent a single average pattern, a series with the cell turned off, a series with the cell active at 1.25 MHz, and then at 3.30 MHz. In the last three cases, the averages of the calculated wave fronts converge towards the first case of measured aberration.

## 6. Results and discussion

Reduction of the speckle noise in our method is achieved in all subjects. This is due to the induced spreading of the incident light beam across the cornea which then forms an extended spot on the retina. The incoming light produces an angularly divergent and frequency-modulated spot that spreads across a somewhat wider region than a tightly focused spot, encompassing in this way a greater number of the highly scattering elements of the retina, such as photoreceptors and blood vessels. This, in effect, results in the averaging of the reflected and scattered light and consequently in the significant reduction of speckle noise and subsequent uniformity of the HS pattern. Lenslet foci, single and groups, which were strongly scintillating, now had nearly equal and smooth intensity distribution (Fig. 5). The increased quality, homogeneity and spatial resolution owing to the smoothing improved the signal to noise ratio of the HS pattern. We also noticed no discrepancy between operating the cell in $x$- or $y$-only orders of diffraction, since the HS lenslets could not resolve the extended retinal spot, and the effect was similar for all resonant frequencies.

We also noticed a slight drop in intensity and contrast when passing through resonance, due to three effects: (a) The retinal focus is less concentrated, but the total scattered light is nearly the same, (b) the higher diffracted orders might miss the edge of the pupil and less light will be admitted into the eye, and (c) different division of power between the orders. The effect of speckle reduction is instantaneous and reproducible by simply turning the signal generator on and off, on demand, when acquiring an image (Fig 5).

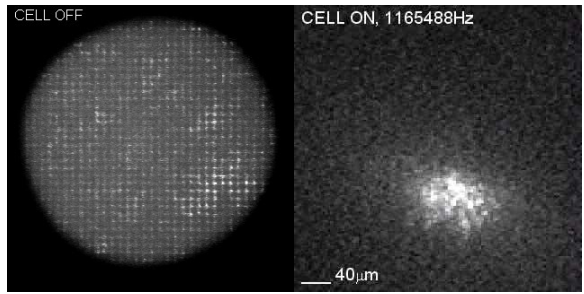

Fig. 5. HS image (left, 1.5 MB, http://physics.technion.ac.il/~eribak/HSDespeckle.avi) and retina focal spot (right, 2.2 MB, http://physics.technion.ac.il/~eribak/RetinalSpot.avi) movies to illustrate the effect of speckle reduction and retinal spot response under the application of the acoustic method. Notice the large number of resonances seen at the retina as the frequency is tuned. The subjective spot image was extremely similar to this double pass spot appearance.

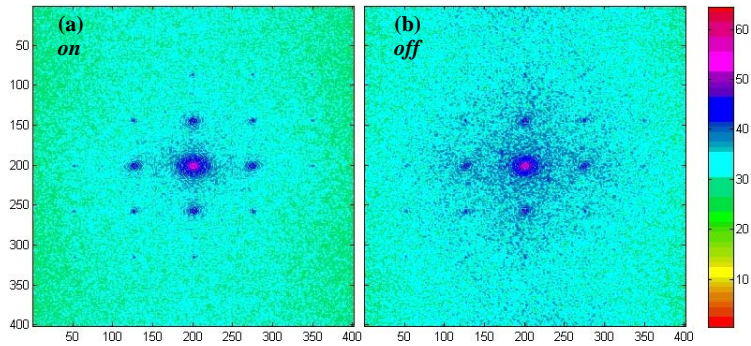

Fig. 6. The power spectrum of a HS pattern obtained with (a) the acoustic cell on, and (b) off (logarithmic scale). While the signal is fully conserved near the side lobes, the background noise is considerably less when the cell is on.

As a typical example, with the acoustic cell operating in a rather strong resonance near 1.2 MHz, we found that the laser spot was elongated to spread approximately 410 µm across the cornea in the *y* direction, namely five or six diffraction orders. As the eye was focused at infinity (gazing at the laser or a target) and the cell was only 40 cm away, the spot extended by about 40µm on the retina. This spread, combined with the angle-of-arrival spread, was found to be sufficient in smoothing the HS pattern and improving the spatial resolution and accuracy of the wave-front. Fig. 6 shows the Fourier transform of one of the HS patterns, with the cell resonating and with the cell off. It is apparent that the signal components in Fourier space are present and are identical for both cases; indicating again that this method does not affect the HS, and therefore the reconstruction of the wave-front. At the same time, the level of background noise drops appreciably when the cell is running. It illustrates that when the cell is operated the noise power is halved and thus the signal to noise ratio is improved, as the HS harmonics are distinctively clearer with respect to the background. The overall standard deviation among the twelve frames, each 768×568×8-bit pixels in size, is $7.70\times10^{-4}$, dropping off by application of the acoustic cell to $4.14\times10^{-4}$. Similar improvement was achieved on the other subjects. Fourier demodulation and band-pass filtering of the HS pattern [8] reduce much of the inherent noise anyway. Thus the benefit is in more efficient usage of the available light, as less of the signal is lost due to scattering into higher frequencies and to camera saturation. In the traditional method of HS analysis, performed by centroiding each focus of the lenslets and fitting a wave-front to the centroid shifts, speckle noise is spread at all frequencies and has a stronger effect. Thus the gain from this acoustic reduction of the speckle noise can be even higher. Using the HS images acquired with the acoustic cell in operation, we proceeded to reconstruct the associated wave-front.

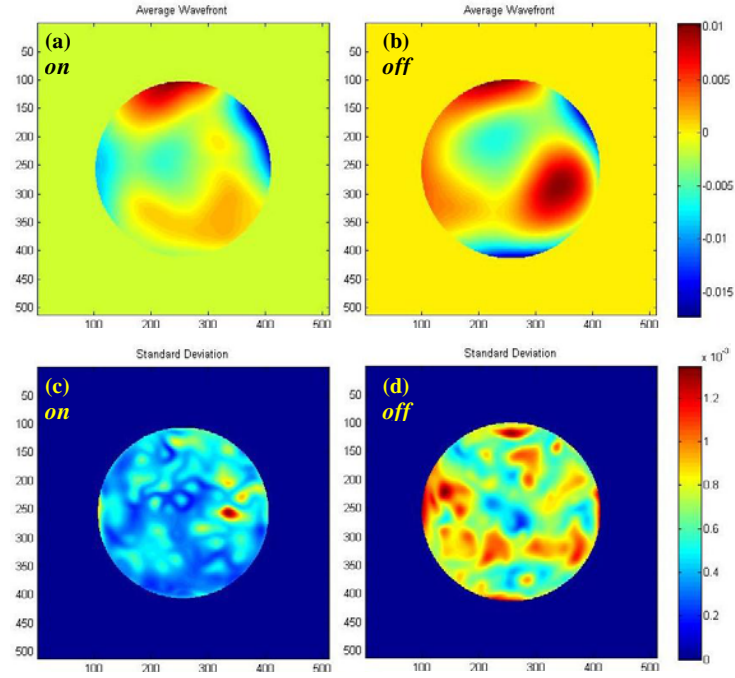

Fig. 7. Average wave fronts and errors [mm], of one subject as reconstructed using the Fourier method from the HS data. (a) and (b) show the average wave front with the acoustic cell on and off respectively, whereas (c) and (d) show the associated standard deviations again with the cell on and off. Accommodation and gaze direction changed slightly between images.

To further illustrate the benefits of this method we also show the average of the twelve reconstructed wave-fronts and their standard deviations when the cell is operating and when it is turned off (Fig. 7). The 'on' wave-front appears to have much fewer features associated with noise, while retaining the same features in both cases.

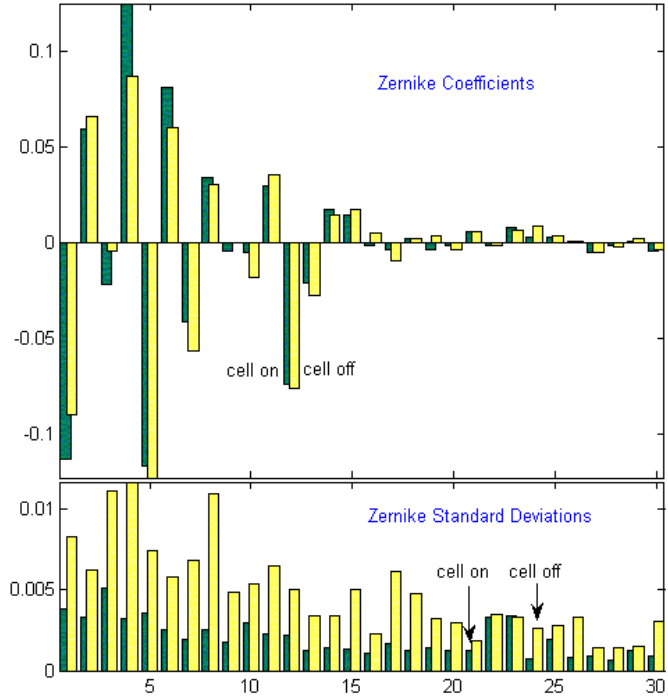

Fig. 8. Zernike coefficients of the same eye (top) and their distributions (bottom) [mm]. The eye was not paralysed, and the coefficients remained nearly the same when the cell was off (yellow bars) and on (green bars). At the same time, the corresponding standard deviations dropped significantly with the application of the cell.

We also decomposed the individual wave-fronts into Zernike coefficients, to compare their distribution with the cell turned on and off. The aberrations have changed between and during these series of images, mostly in direction and defocus (Fig. 7). The speckle content and saturation was different, and as a result the variance between the wave fronts within each group is different. It is clear that the errors are much smaller for the 'on' group (Fig. 8).

## 7. Conclusions

In summary we have presented a method to reduce the speckle pattern and make it more uniform. There are two benefits, the first and foremost reducing the speckle noise and variability of the measurements. Another advantage is that the speckles do not saturate the detector and do not distort the measurement. Since the light level is more uniform, the dynamic range of the detector can be extended and the measurements become more accurate (or a simpler detector can be used). The result is that the reconstructed wave-fronts are more precise and repeatable. The method offers simplicity in arrangement since the acoustic cell is placed in tandem with the light source and does not interfere with the wave front sensor. While the smoothing can be performed with any light source, such as collimated superluminescent diodes or white light, it is now possible to use the more coherent laser, which is easier to align and point and more power-efficient. Other wave front sensors such as the pyramid [12, 13] or curvature [14] also requiring speckle-free reference sources, can gain from this solution. Finally, the application is valid for other speckle-limited sensing, such as scanning-laser ophthalmoscopes, tissue microscopy, and non-biological sensing.

## 8. Acknowledgments

Parts of this work were performed under the funding of the Israeli Ministry of Science and the French-Israeli science fund. Engineering assistance from Shamir Optical Industry is greatly appreciated.